\documentclass[aps,pra,reprint,superscriptaddress,twocolumn,notitlepage,longbibliography]{revtex4-1}

\usepackage{times,bbm}

\usepackage[usenames,dvipsnames]{xcolor}
\usepackage{mathtools}
\usepackage{tikz}  
\usetikzlibrary{arrows,shapes,positioning,shadows,backgrounds,fit}

\usepackage[english]{babel}
\usepackage{graphicx, dcolumn, bm, color, amsmath,amsthm, relsize, amsmath, dsfont, mathrsfs, empheq, verbatim, upgreek, etoolbox,braket}
\usepackage[caption = false]{subfig}

\usepackage[colorlinks=true,linkcolor=teal,urlcolor=teal,citecolor=teal]{hyperref}
\usepackage{graphics}
\usepackage{bm}
\usepackage{color}
\usepackage{amscd}
\usepackage{amsfonts}
\usepackage{amssymb}
\usepackage{graphicx}
\usepackage{tabularx}

\newcommand{\Ham}{\mathcal{H}}

\newcommand{\nbar}{\overline{n}}
\newcommand{\ketbra}[2]{|#1\rangle\langle#2|}
\newcommand{\blah}{blah\\blah\\blah\\blah\\blah\\blah\\blah\\blah.}

\usepackage{amsthm}
\theoremstyle{definition}

\newtheorem*{theorem*}{Theorem}

\newtheorem*{corollary*}{Corollary}

\newtheorem*{lemma*}{Lemma}

\begin{document}

\title{Stability of emergent time periodicity in a few-body interacting system} 

\author{Steve Campbell}
\affiliation{School of Physics, University College Dublin, Belfield, Dublin 4, Ireland}
\affiliation{Centre for Quantum Engineering, Science, and Technology, University College Dublin, Dublin 4, Ireland}
\affiliation{Dahlem Center for Complex Quantum Systems, Freie Universit\"{a}t Berlin, 14195 Berlin, Germany}

\author{Jens Eisert}
\affiliation{Dahlem Center for Complex Quantum Systems,
Freie Universit\"{a}t Berlin, 14195 Berlin, Germany}

\author{Giacomo Guarnieri}
\affiliation{Department of Physics, University of Pavia, Via Bassi 6, 27100, Pavia, Italy} 
\affiliation{Dahlem Center for Complex Quantum Systems, Freie Universit\"{a}t Berlin, 14195 Berlin, Germany}

\begin{abstract}
We examine the onset and resilience of emergent time periodicity in a few-body all-to-all interacting Lipkin-Meshkov-Glick model, where one of the constituents is locally in contact with a thermal bath. Employing both a collision model framework and a suitable time-continuous description, we show that stable time-periodic behavior can only be exhibited when the bath acts as a purely dissipative channel. We assess the role that the microscopic interactions within the system play, establishing that for the all-to-all model the introduction of temperature leads to a melting of the emergent time periodicity, in contrast to stable long-time behavior which can be maintained for nearest neighbor $XXZ$ type interactions.
\end{abstract}
\maketitle

\section{Introduction}
Any realistic physical system unavoidably interacts with a surrounding environment to some extent. The resulting non-unitary dynamics typically leads open quantum systems to equilibrate. Pursuing this line of inquiry has led to remarkable insights into a wealth of phenomena, including notions of decoherence and dissipation~\cite{Alicki2007, Breuer2002}, the emergence of classical objectivity in a quantum prescription~\cite{ZurekRMP, RyanJPhycComms}, the appearance of decoherence free subspaces~\cite{DFSreview} and the quantum Zeno effect~\cite{Zeno}, to name but a few. Equilibration, however, typically entails the reduction or even loss of genuine quantum properties, such as coherence and locally accessible entanglement that would guarantee an edge with respect to classical technologies in several applications ranging from quantum metrology~\cite{Smirne2016a} to quantum simulation \cite{CiracZollerSimulation} and (near-term) quantum computing~\cite{Arute2019,SupremacyReview,Variational}.

A particularly interesting class of systems that seem to buck this trend are those displaying non-stationary behaviors in their long time dynamics~\cite{BucaNatComms, BucaSciPost2022}. Such systems, sometimes referred to as {\it ``dissipative time crystals"}, correspond to situations where a many-body system in contact with an environment fails to reach equilibrium and instead exhibits emergent time periodicity. While the study of this interesting phenomenon has its roots in the original proposal for the spontaneous breaking of time translation symmetry~\cite{Wilczek2012} which is now characteristic of \emph{discrete} time crystals, it is important to stress that the time-periodic behaviour exhibited by these so-called dissipative time-crystals arises from a  different mechanism. In fact, dissipative time crystals lack any inherent underlying time-symmetry, relying instead on the onset of oscillations between degenerate asymptotic states of the dissipative quantum channel~\cite{riera2020time}. Notwithstanding such subtle differences, both discrete and dissipative time-crystals have since spurred a wide ranging literature~\cite{DTCReview, riera2020time, JinPRA2023, MoessnerPRB2017, KhemaniPRXQ,ReviewTimeCrystals} showing that time crystalline behavior can be exhibited in a variety of settings, including boundary~\cite{iemini2018boundary}, Floquet~\cite{FloquetTC}, Stark~\cite{EisertPRB}, and higher dimensional~\cite{Augustine2DTC} time crystals. Common to all these realisations is that emergent time-periodic behavior necessarily requires some driving, either in the form of an explicit time-periodic kick, as is the case of Floquet time crystals, or by explicitly coupling to a bath~\cite{BucaNatComms, BucaSciPost2022}. While the prescription of a dissipative time crystal gives rise to a wealth of specific schemes, their actual stability is in many instances far from clear~\cite{ReviewTimeCrystals}, with few examples with proven stability being known~\cite{PhysRevB.108.L180201,GiacomoPRA}.

Within the plethora of techniques to model an open system dynamics, collision models (also known as repeated interaction schemes) present a particularly versatile framework~\cite{Ciccarello_2022, CampbellEPLReview, RauPR}, one which may also allow to formulate quantitative statements more easily. By giving partial access to the relevant environmental degrees of freedom, they allow for a more precise microscopic description of the dynamics to be derived. They have found useful application in the study of time crystalline behavior, including assessing the thermodynamics of boundary time crystals~\cite{GabrieleArXiv}, and in establishing emergent time periodicity from an otherwise fundamentally random dynamics~\cite{GiacomoPRA}, which will be the starting point of our work. 

In this work, we study the stability of the emergent time periodic behavior established in Ref.~\cite{GiacomoPRA} to both changes in the microscopic description of the internal interactions and the resilience (or lack thereof) to thermal effects. As main results we show that the establishment of a decoherence free subspace leads to a time periodic behavior. However, its stability is significantly impacted by the precise details of thee generalised dynamical symmetry that supports the oscillations. In particular, we show that a fully connected system modelled by the Lipkin-Meshkov-Glick Hamiltonian satisfies the conditions necessary to be a dissipative time crystal~\cite{BucaNatComms, GiacomoPRA} only in the case of a zero temperature environment. For finite temperatures, we show the time crystal melts and relate this to the specific characteristics of the generalised dynamical symmetry. Finally, we revisit the $XXZ$ model studied previously by some of us~\cite{GiacomoPRA} and show that, remarkably, in a small system of $N\!=\!4$ spins such a few-body system is sufficient to host protected symmetries that are fundamentally stable to arbitrarily strong thermal effects.

\section{Emergent time periodicity from collisional dynamics}
Collision models~\cite{PhysRevA.80.022334,Ciccarello_2022,CampbellEPLReview,RauPR,PhysRevLett.95.110503,cMPS2} have emerged as a versatile tool for simulating and understanding a range of open system dynamics (and of quantum fields~\cite{cMPS1,cMPS2}). We model the environment as a large, in principle infinite, collection of identical copies of an auxiliary quantum system, $A$. An effective open system dynamics for the system $S$ is achieved by allowing its free evolution to be punctuated by interactions (collisions) with a single copy of the auxiliary system for a given interaction time $\tau$. While at first glance such a model may appear to be an over simplification, a remarkably broad range of physically relevant settings can be modeled using this approach~\cite{Ciccarello_2022}. In particular, any Markovian dynamics, including a system in contact with a large thermal bath as will be the focus of the present work, can be simulated by a collision model~\cite{DeChiaraPRL2021}. In what follows, we will assume that the system is a many-body quantum system composed of many constituents or `sites', and that the collisions take place only between a single fixed site (without loss of generality hereafter denoted with $q_1)$ and the $n$-th auxiliary system, governed by a simple exchange interaction Hamiltonian
\begin{equation}
\label{interactionHam}
\hat{H}_{q_1,A_n}= \frac{1}{2} \left(\hat\sigma_x^{(1)} \otimes \hat\sigma_x^{(A_n)} + \hat\sigma_y^{(1)} \otimes \hat\sigma_y^{(A_n)}\right). 
\end{equation}
Tracing out the auxiliary degrees of freedom after each interaction thus leads to an effective open system dynamics for $S$.

We briefly recapitulate the main result of Ref.~\cite{GiacomoPRA} where it has been shown that time periodicity can emerge from an otherwise random dynamics. We will assume that the waiting time between two subsequent collisions is random, following a probability distribution $p(\theta)\sim e^{-\gamma \theta}$ with $\gamma\tau \ll 1$ in order to guarantee that the system always interacts with only one auxiliary system at a time. It is important to stress that the specific choice of $p(\theta)$ does not impact our main results and any arrival scheme compatible with the above constraint, including fully deterministic ones do not affect our findings. The dynamics after $n$ collisions will result in the state
\begin{equation}\label{totalmap}
    \hat\rho_{S,n} = \text{Tr}_{A_n, \dots A_1}\left[ \underbrace{\mathcal{U}_{S,\theta_n,A_n}\circ\ldots\circ\mathcal{U}_{S,\theta_{2},A_{2}}\circ\mathcal{U}_{S,\theta_{1},A_{1}}[}_{\text{n times}}\hat\rho_{S,0}] \right],
\end{equation}
where $\theta_1,\ldots,\theta_n$ are $n$ possible outcomes of the random variable $\theta$, $\mathcal{U}_{S,\theta_j,A_j}[\cdot] = \hat{U}_{q_1,A_j}\hat{U}_S(\theta_j)\cdot\hat{U}^{\dagger}_S(\theta_j)\hat{U}^{\dagger}_{q_1,A_j}$ gives a period of free evolution followed the collision of duration $\tau$ governed by 
\begin{equation}
\hat{U}_{q_1,A_j} = \exp\left[-i \tau \left(\hat{\mathcal{H}}_S+\hat{H}_{q_1,A_n}\right)\right].
\end{equation}

We assume the auxiliary qubits of the collisional environment have free Hamiltonian given by $\hat{H}_A=-B \hat\sigma_z$ and that they are initialized in thermal states at inverse temperature $\beta=1/T$ and are therefore described by the density matrix
\begin{equation}
\label{ancilla_state}
\hat\rho_A = \frac{1}{2} \begin{pmatrix}
		1-\tanh \left( \tfrac{\beta B}{2} \right) & 0 \\
		0 & 1+\tanh \left( \tfrac{\beta B}{2} \right)
\end{pmatrix},
\end{equation}
which allows to easily consider the impact of a thermal bath on the ensuing system dynamics.

Within this framework, it has been proven in Ref.~\cite{GiacomoPRA} that, if the system supports the existence of a so-called \textit{dynamical symmetry}, which we will rigorously define shortly in Eq.~\eqref{dynamicalSymmetry}, then any operator having non-zero overlap with it would be guaranteed to display robust and stationary oscillations. To be more precise, let us first of all introduce the reduced quantum channel, i.e., a
\emph{completely positive and trace-preserving} (CPTP) map, describing a single collision
\begin{equation}
    \Lambda_{\tau,\theta} [\cdot] = \mathrm{Tr}_A\left[ \mathcal{U}_{S,\theta_j,A_j}[\cdot]\right] = \sum_k \hat{\Omega}_{k}(\tau) \cdot \hat{\Omega}_{k}^{\dagger}(\tau), 
\end{equation}
where we have introduced the Kraus 
operators 
\begin{equation}
\hat{\Omega}_k(\tau) := \hat{\Omega}_{\alpha\beta,\tau} = \sqrt{p_\alpha} \bra{\beta} \hat{U}_{q_1,A_j}\hat{U}_S(\theta_j) \ket{\alpha}
\end{equation}
and $\hat\rho_A = \sum_\alpha p_\alpha\ket{\alpha}\bra{\alpha}$. It is easy to see that by construction, these Kraus operators give rise to a trace-preserving map.

Owing to being CPTP, all the eigenvalues of $\Lambda_{\tau,\theta}$ lie either within the unit circle in the complex plane (\textit{decaying states}) or on its boundaries (\textit{peripheral spectrum}); in 
particular (with a slight abuse of notation motivated by consistency with Sec.~\ref{BucaConds}) we will denote with $\hat\rho_{\infty}$, the eigenstate corresponding to eigenvalue $1$ (which always exists because of the trace preserving nature of the quantum channel), 
i.e., $\Lambda_{\tau,\theta}\left[\hat\rho_{\infty}\right] = \hat\rho_{\infty}$. 
Following 
Refs.~\cite{BucaNatComms,medenjak2020rigorous,chinzei2020time}, we define a dynamical symmetry as an operator $\mathcal{A}$ satisfying the following two properties 
\begin{align}
\label{dynamicalSymmetry}
&\mathrm{\mathbf{(i)}}\, \left[\hat{\Ham}_S,\hat{\mathcal{A}}\right] = -\lambda \,\hat{\mathcal{A}},\qquad \lambda\in\mathbb{R}\notag,\\
&\mathrm{\mathbf{(ii)}}\, \left[\hat{\Omega}_{k}(\tau), \hat{\mathcal{A}}\right] \hat\rho_{\infty} = 
0,\qquad\forall k,\tau,
\end{align}
namely an eigen-operator of the system Hamiltonian [(i)] which also spans a subspace which is left invariant by the collisions [(ii)]. 
In this case, it can be demonstrated~\cite{GiacomoPRA} that the evolution of the operator $\hat{\mathcal{A}}\,\hat\rho_{\infty}$ becomes a clean oscillation    
\begin{equation}\label{theorem:Evol}
\Lambda_{\tau,\theta}[\hat{\mathcal{A}}\,\rho_{\infty}] = e^{-i\lambda (\tau+\theta)} \,\hat{\mathcal{A}}\,\hat\rho_{\infty},
\end{equation}
at frequency $\lambda$ which depends solely on the spectrum of the system's Hamiltonian and not on any underlying periodic driving. We refer the interested reader to Ref.~\cite{GiacomoPRA} for a more complete derivation of the above.

\begin{figure*}[t]
(a)\hskip0.65\columnwidth (b)\hskip0.65\columnwidth (c) \\
\includegraphics[width=0.65\columnwidth]{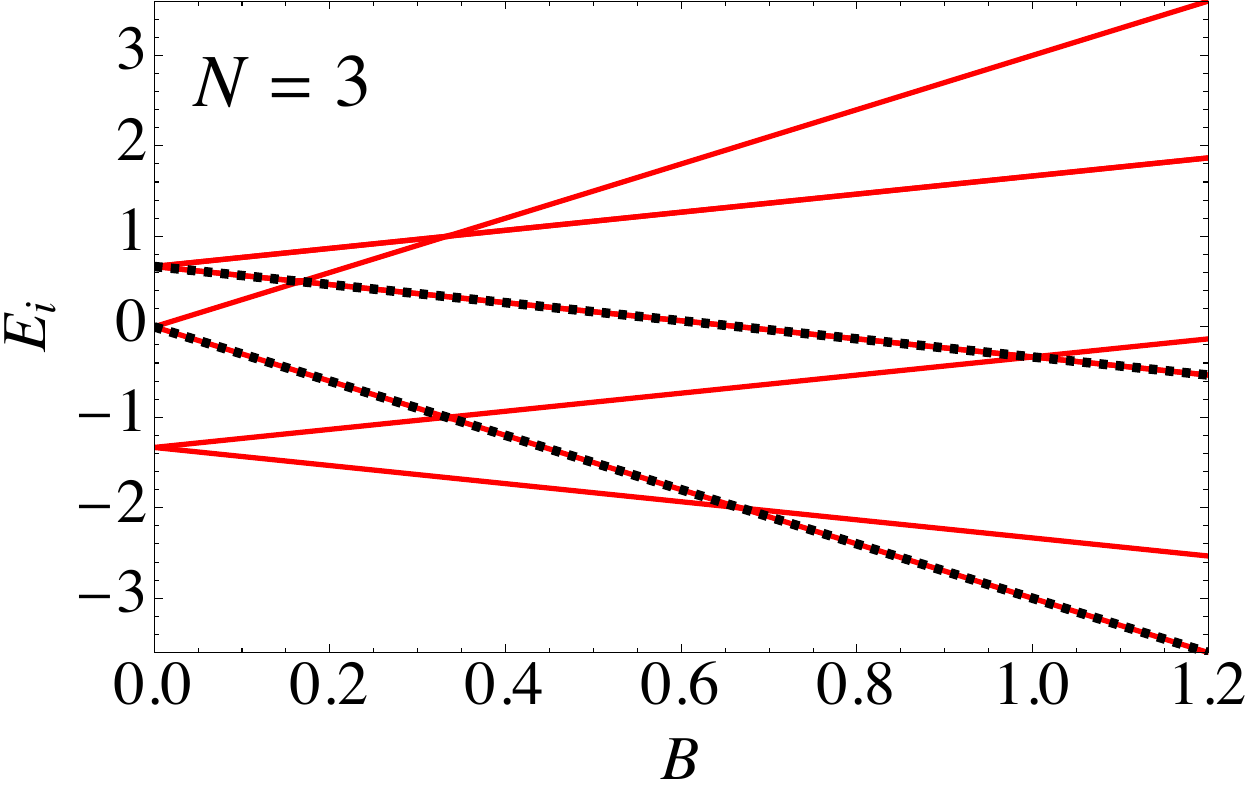}~~~~~\includegraphics[width=0.65\columnwidth]{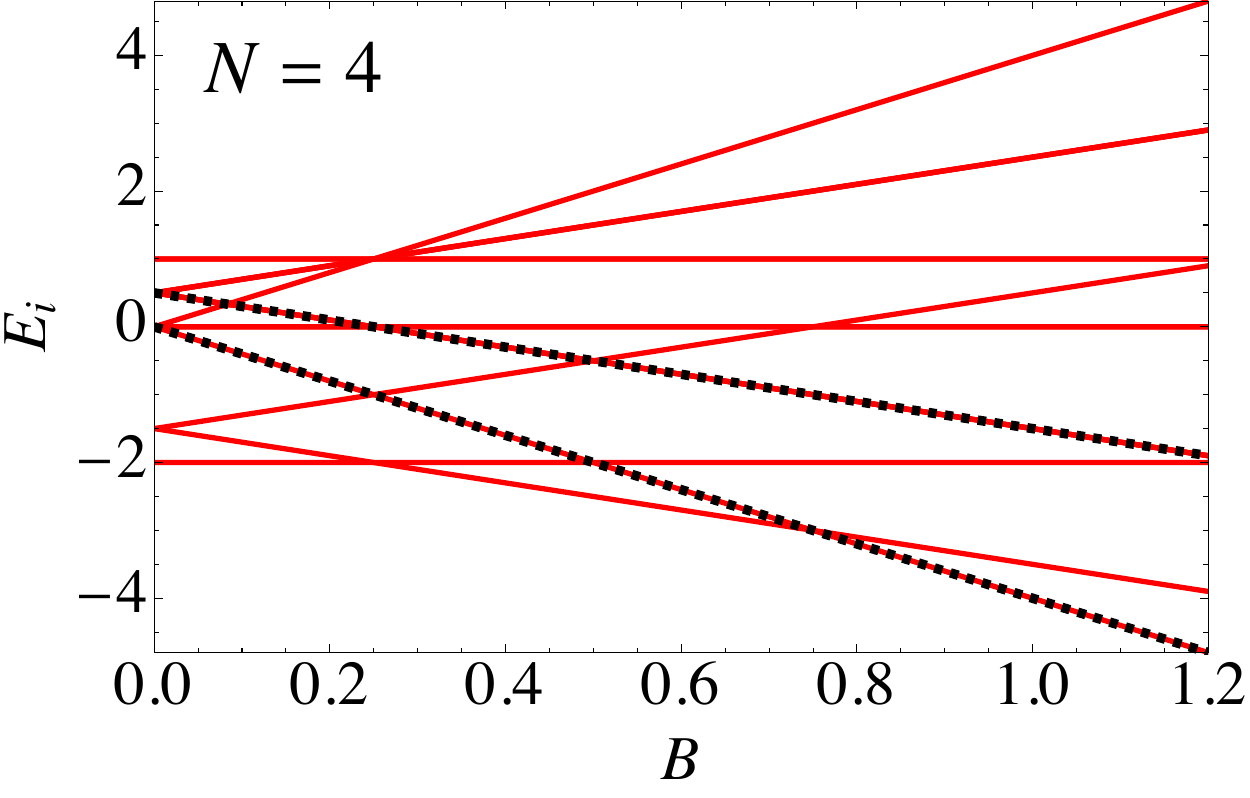}~~~~~\includegraphics[width=0.65\columnwidth]{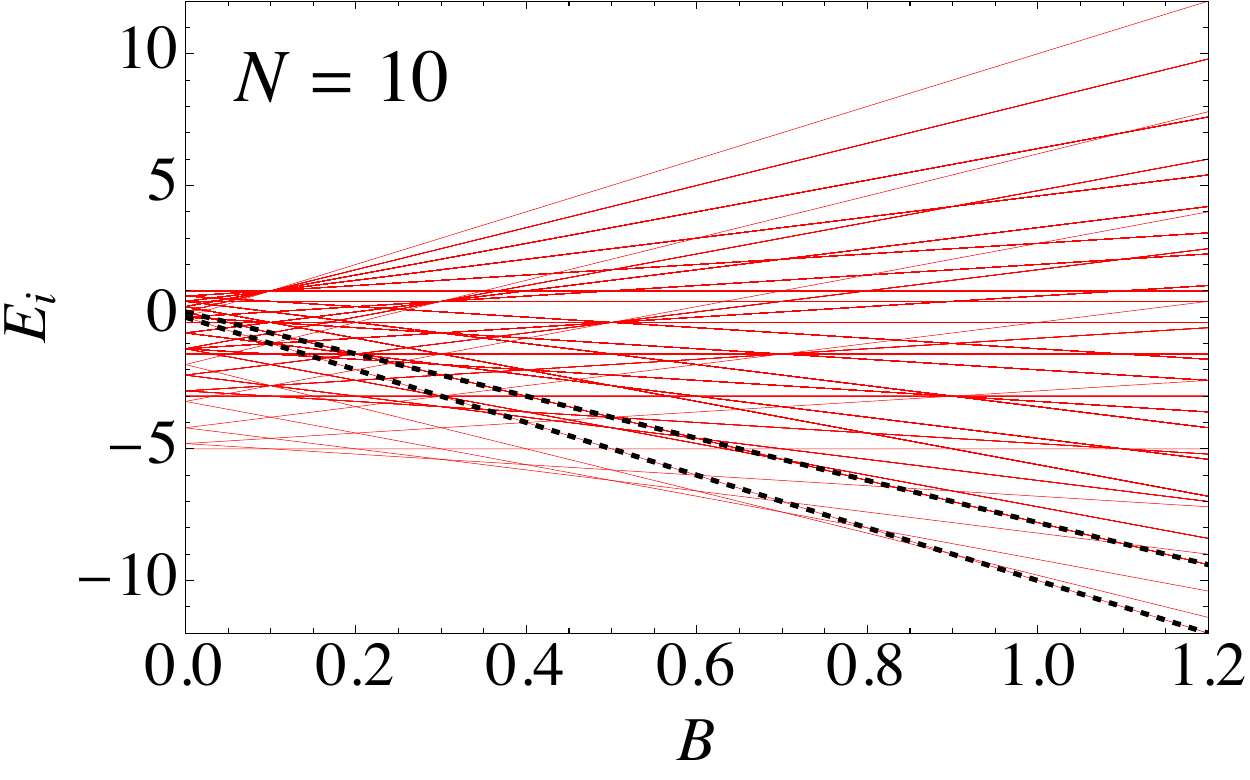}
\caption{Energy spectra for the isotropic LMG model, Eq.~\eqref{LMG_Ham}. We show $N=$3, 4, and 10 respectively. The dashed lines on each plot indicate the energy eigenstates that form the dynamical symmetry of the model for the zero temperature environment case.}
\label{fig1_spectra}
\end{figure*}

While this statement holds true for any dynamics described by a CPTP map, the specific structure 
of collision models plays a crucial role in autonomously selecting this periodic dynamics, getting rid of any other subspace.
Exploiting the above-mentioned fact that all quantum channels $\tilde{\Lambda}_{\tau,\theta}$ share the same peripheral spectrum, taking the limit of many collisions $n \gg 1$ automatically implies that all decaying states (whose eigenvalues lie within the unit circle and thus are in modulus smaller than 1) get shrank down to zero, thus leaving only the oscillating states in the peripheral spectrum $\hat{\mathcal{A}}\,\hat\rho_{\infty}$ untouched. As a consequence, the expectation value of any system observable $\hat{O}$ that has a non-zero initial overlap with the dynamical symmetries of the model in the Hilbert-Schmidt sense, i.e., $ \mathrm{Tr} [\mathcal{A}\hat\rho_{\infty}\hat\rho_S^{(0)}] \neq 0$, after sufficiently many collisions will display oscillations at frequency $\lambda$.

\section{Results}
We consider $N$ qubits labeled $q_i$ with $i=1,\dots, N$ described by the isotropic all-to-all interacting model, the \emph{Lipkin-Mashkov-Glick} (LMG) model, given by the Hamiltonian
\begin{equation}
    \label{LMG_Ham}
    \hat{\Ham} = - \frac{J}{N} \sum_{i<j} \left( \hat{\sigma}^{(i)}_x \hat{\sigma}^{(j)}_x +\hat{\sigma}^{(i)}_y \hat{\sigma}^{(j)}_y \right) - B\sum_{i=1}^N \hat{\sigma}_z^{(i)},
\end{equation}
where $\hat\sigma_\alpha$ are the usual Pauli matrices, $B$ is an on-site magnetic field and $J$ is the coupling~\cite{LMG1, *LMG2, *LMG3, LMG4}. This and closely related models have been extensively studied, particularly in the thermodynamic limit where they display a range of interesting critical phenomena~\cite{LMGQPTs, ESQPTReview}. The high degree of symmetry endows the model with an interesting energy spectrum which we show examples of in Fig.~\ref{fig1_spectra} for various sized systems. The cascade of energy level crossings for increasing field strength is characteristic of such isotropic interacting models. In what follows, we consider a similar setting to Ref.~\cite{GiacomoPRA} where qubit $q_1$ is coupled to the collisional bath and we examine the ability for Eq.~\eqref{LMG_Ham} to admit stable time periodicity when observing any single qubit of the system.

\subsection{Zero temperature analysis}

\label{ZeroTempSection}
We begin by considering a zero-temperature collisional bath, i.e., taking $\beta\to\infty$ in Eq.~\eqref{ancilla_state} and, therefore, $\hat\rho_A = \ketbra{0}{0}$. This drives $q_1$, the qubit in contact with the collisional bath, to its ground state in the long time limit. For $N=2$, Eq.~\eqref{LMG_Ham} does not support any stable oscillatory behavior, with both qubits ultimately being driven to their local ground states~\cite{MariaPRA}. In contrast, for $N\!\geq\!3$ the spectrum of the system is sufficiently rich that it can support stable oscillations emerging from the otherwise intrinsically random dynamics of the collision model since the additional qubits in the system allows for decoherence free subspaces to be supported~\cite{DFSreview}. These subspaces live between specific energy eigenstates of the system Hamiltonian. The bath locally interacting with only the first qubit in the system is sufficient to open a dissipative channel which can ``purge" the system of unwanted eigenstates and thus this setting establishes the conditions necessary for dissipative system to exhibit an emergent time-periodic behavior~\cite{BucaNatComms}. 

\begin{figure}[t!]
(a)\\
\includegraphics[width=0.9\columnwidth]{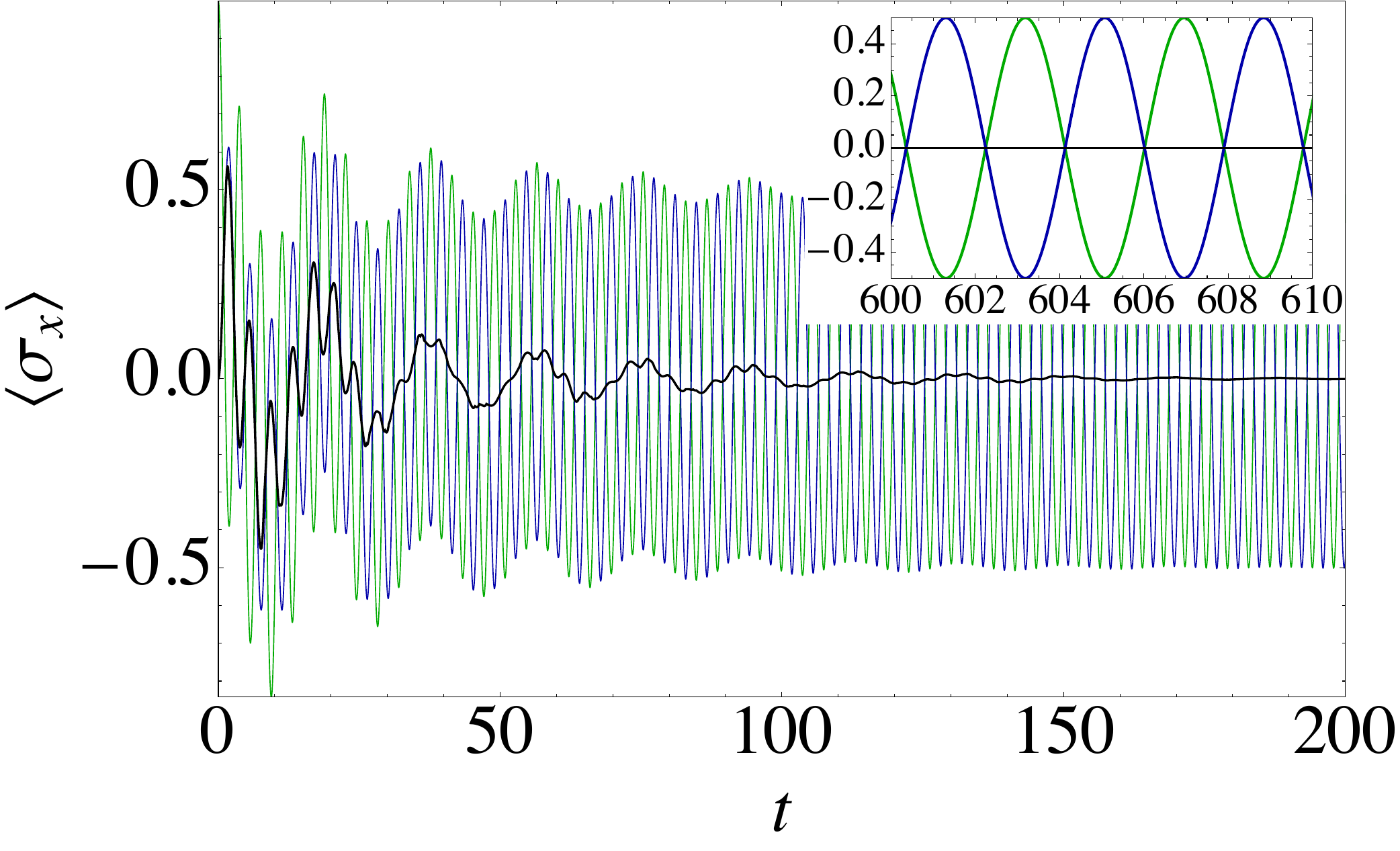}\\
(b)\\
\includegraphics[width=0.9\columnwidth]{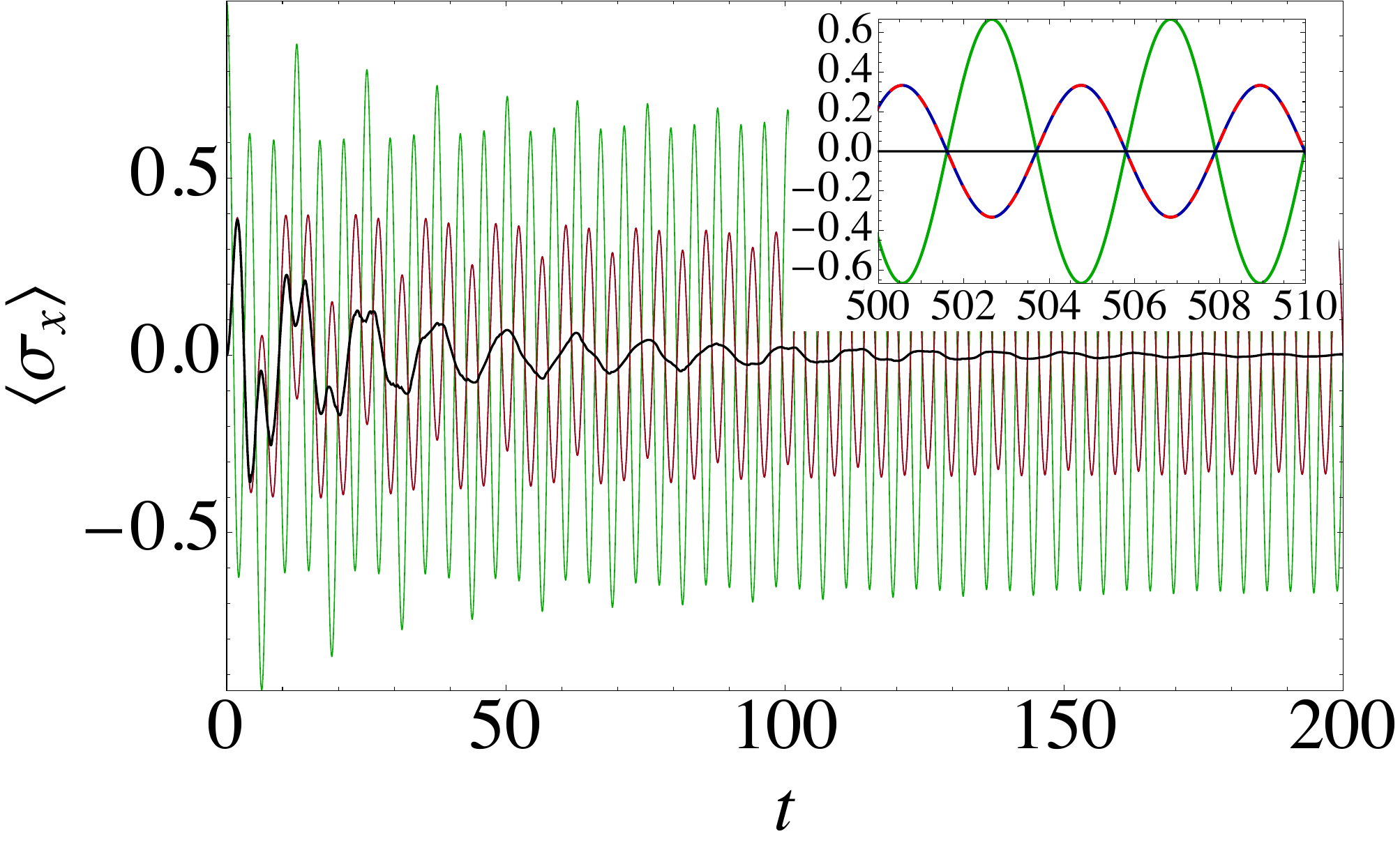}\\
(c)\\
\includegraphics[width=0.83\columnwidth]{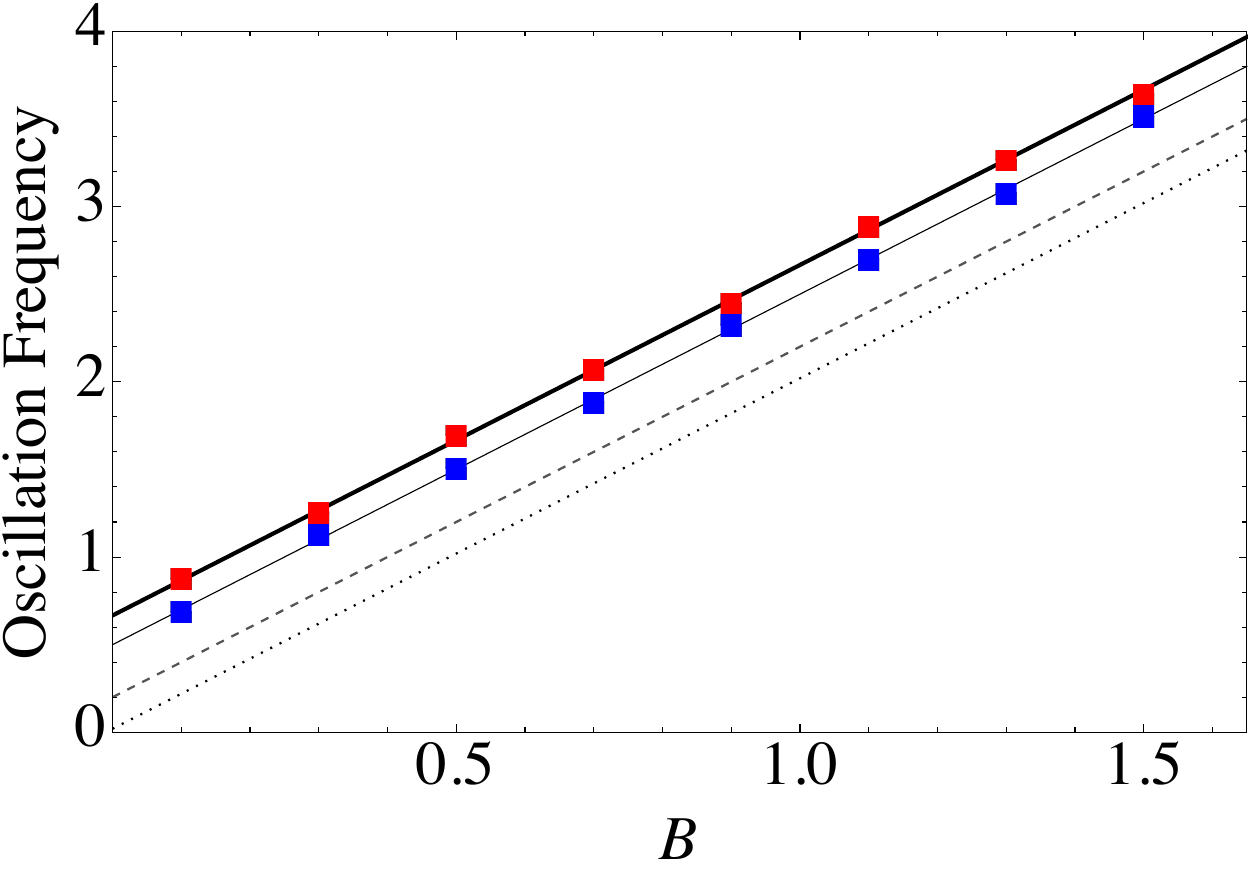}
\caption{We show the behavior for a fixed initial state vector $\ket{\Psi_0}$ given by Eq.~\eqref{initialstate}. We choose $B/J=0.5$, consider 400 collisions with collision duration 0.5 and an exponential distribution for the random arrival times with $\gamma=1$. In all plots $\langle\hat\sigma_x^{(1)}\rangle$ [qubit that interacts with collisional bath] is shown in black. (a) For $N=3$ we see $\langle\hat\sigma_x^{(2)}\rangle$ and $\langle\hat\sigma_x^{(3)}\rangle$ are exactly anti-periodic. (b) For $N=4$ we see $\langle\hat\sigma_x^{(2)}\rangle$ has the largest amplitude while $\langle\hat\sigma_x^{(3)}\rangle$ and $\langle\hat\sigma_x^{(4)}\rangle$ exactly coincide. (c) We show the oscillation frequency in the long time limit for the zero temperature collisional bath case, the line is given according to Eq.~\eqref{OscFreq} for $N=3$, 4, 10, and 100 [top to bottom] while the points for $N=3$ and 4 are determined directly from numerical simulations.}
\label{fig2_zerotemp}
\end{figure}
To make things concrete, in Fig.~\ref{fig2_zerotemp} we examine the case of $N=3$ and 4 explicitly. We fix the initial state vector to be 
\begin{equation}
\label{initialstate}
\ket{\Psi_0} = \ket{0}^{(1)} \otimes \ket{+}^{(2)} \otimes \left( \bigotimes_{j=3}^N \ket{0}^{(j)} \right),
\end{equation}
and (arbitrarily) choose $B=0.5$. For both system sizes we find stable oscillations are established in all of the qubits that are not directly interacting with the collisional bath, while $q_1$ is clearly driven to an equilibrium state at zero temperature. We remark that only the amplitude of the observed oscillations are impacted for any other choice of initial state, including fully random, that has a non-zero overlap with the dynamical symmetry~\cite{GiacomoPRA}. Since the period of oscillations is directly related to the relevant spectral gap in the system Hamiltonian, i.e., it 
is the energy difference between the  states supporting the decoherence free subspace, we can readily identify which energy eigenvalues and states that the system will oscillate between in the long time limit, which are shown in Fig.~\ref{fig1_spectra} by the dotted lines, and are given by
\begin{equation}
\label{eigenvalues}
E_\nu = -N B, \qquad \text{and} 
\qquad E_\mu = \frac{2}{N}-(N-2)B.
\end{equation}
The eigenvector corresponding to $E_\nu$ is simply $\ket{\psi_\nu}=\prod_{i=1}^{N} \ket{0}^{(i)}$, while $E_\mu$ is an eigenvalue with $(N-1)$ fold degeneracy. We 
find that the system exhibits clean, persistent oscillations with a frequency given by
\begin{equation}
\label{OscFreq}
\lambda = E_\mu - E_\nu = \frac{2}{N}+2B.
\end{equation}
Immediately we see that in the limit of $N\to \infty$ the frequency tends to that of a freely (Rabi) oscillating qubit as a consequence of the vanishing interaction between the observed qubit and $q_1$ in Eq.~\eqref{LMG_Ham}. 

Considering $N=3$ in Fig.~\ref{fig2_zerotemp}(a), we see that $\langle\hat\sigma_x^{(2)}\rangle$ and $\langle\hat\sigma_x^{(3)}\rangle$ oscillate with the same amplitude but perfectly out of phase with each other, indicating that any collective measurement, e.g.,
\begin{equation}
\hat{S}_x=\sum_i \hat\sigma_x^{(i)} 
\end{equation}
will not show an oscillatory behavior due the cancellation between the two subsystem expectation values~\footnote{We remark that other relevant quantities such as the correlation functions $\langle \sigma_x^{(j)} \sigma_x^{(k)} \rangle$ for $N\!=\!3$ and $4$ also do not exhibit periodic non-stationary behavior in the long time limit. This further evidences that the present setting does not support long-range order and therefore is distinct from other notions of time-crystals~\cite{WatanabePRL2015}.}. Nevertheless the individual qubits are stably oscillating with the expected frequency given in Eq.~\eqref{OscFreq}. In Fig.~\ref{fig2_zerotemp}(b) we consider $N=4$. Due to the all-to-all nature of the coupling in Eq.~\eqref{LMG_Ham}, $q_2,~q_3,$ and $q_4$ all equally feel the effect of the collisional bath. The only asymmetry present is at the level of the initial state, Eq.~\eqref{initialstate}. From Fig.~\ref{fig2_zerotemp}(b) we again see that regardless of which of these qubits we choose to measure the system still admits stable oscillations with exactly the expected frequency. However, we now see that the amplitude of the oscillations is impacted. For $q_2$, which 
has been initialised in $\ket{+}$, the amplitudes are larger compared to $q_3$ and $q_4$. 

Finally, in Fig.~\ref{fig2_zerotemp}(c) show explicitly the linear dependence that the observed oscillation frequency on the value of the field, $B$. The points for $N\!=\!3$ and 4 are computed directly by numerically simulating the dynamics and taking the Fourier transform of the long time behavior of $q_2$ to determine the oscillation frequency while the lines correspond to Eq.~\eqref{OscFreq}. We clearly see that despite the model exhibiting energy level crossings as a function of $B$, cfr.\ Fig.~\ref{fig1_spectra} (and a QPT in the thermodynamic limit), these features play no role in the observed oscillatory behavior. As the system is scaled up, the emergent time-periodic behavior is effectively lost and the observed qubit freely oscillates. This indicates the the establishment of a dissipative time-crystal for this model is dependent on the interplay between the (collisional) bath in contact with the system via qubit $q_1$ 
and the remaining qubits, $q_{i\neq1}$, in the system that act as an effective second bath for the observed qubit, ultimately driving the system periodically in a non-trivial manner. 

\subsection{Finite temperature analysis}
\label{FiniteTempSection}
\begin{figure}[t]
(a)\\
\includegraphics[width=0.95\columnwidth]{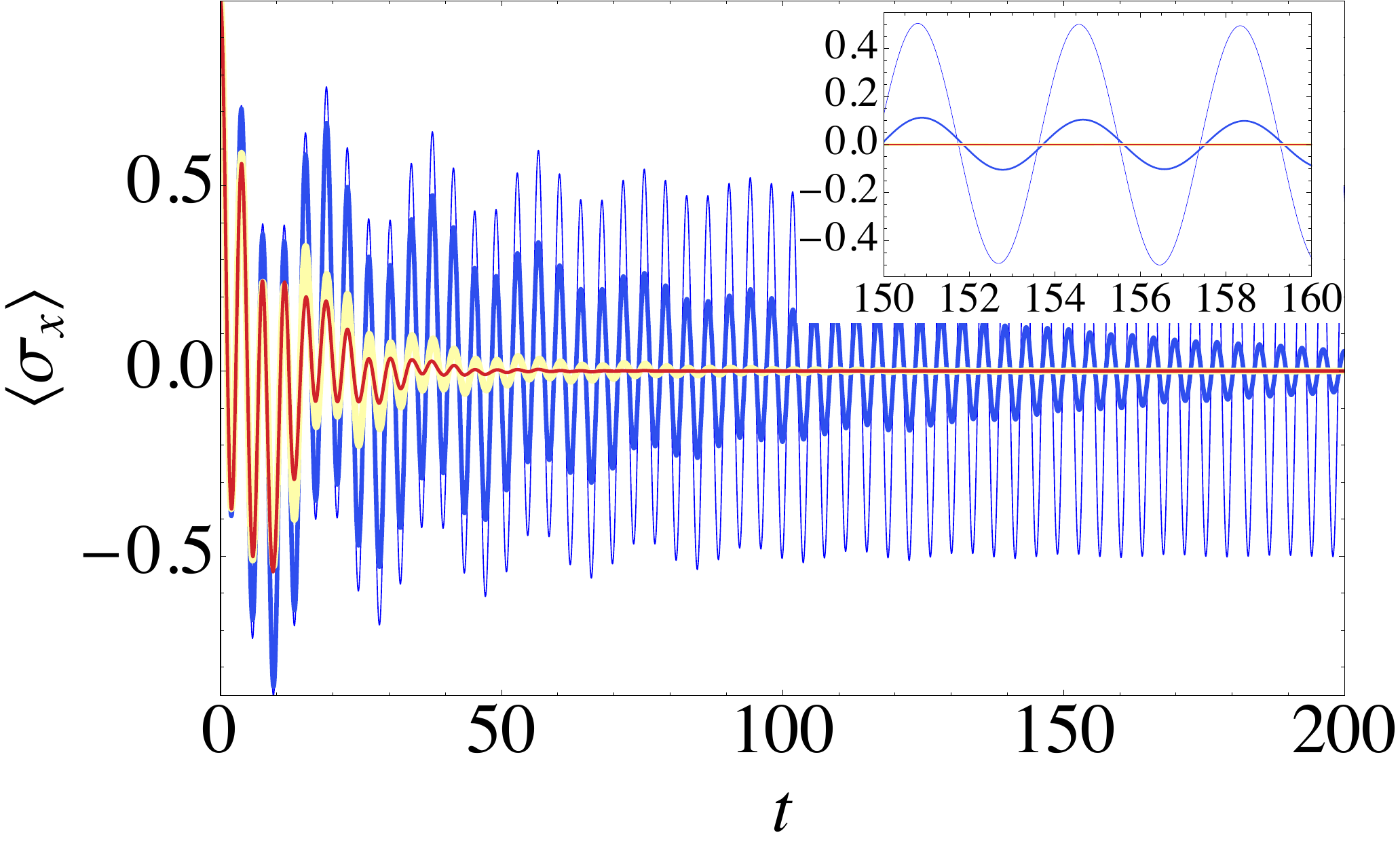}\\
(b)\\
\includegraphics[width=0.9\columnwidth]{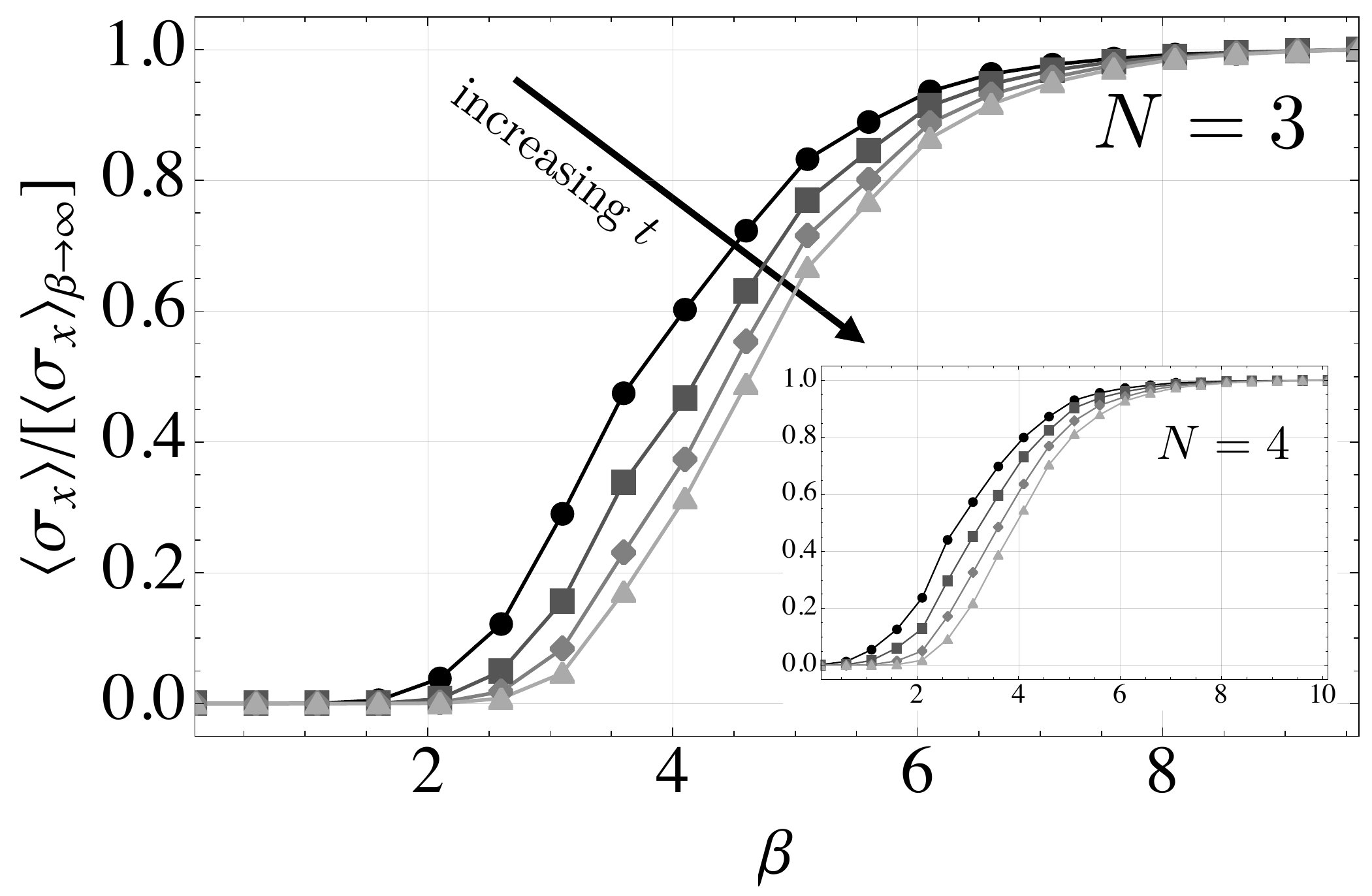}
\caption{(a) Finite temperature $\langle \hat\sigma_x^{(2)} \rangle$ for $N=3$ with same initial state, Eq.~\eqref{initialstate}. We show different values of (inverse) temperature for the auxiliary systems' states corresponding to $\beta=10$ (thin, blue),  2.5 (thick, blue), 1.0 (lighter, yellow) and 0.1 (red) following a heat map color scheme. Inset shows a zoomed in region of time. (b) Decay of oscillations  observed as a function of $\beta$ at different instances of time starting from $t=200$, 300, 400, and 500 from left to right. Inset shows a qualitatively similar behavior holds for increased system size, $N=4$.}
\label{fig3_FiniteTemp}
\end{figure}

From the previous section it is clear that the oscillatory behavior is robust at the level of the Hamiltonian parameters, in particular for finite $N$ it occurs regardless of the specific value of interaction strength and/or applied field strength. We now consider how the nature of the collisional bath impacts the dynamical behavior by demonstrating that the introduction of temperature at the level of the auxiliary qubits making up the collisional bath degrades the oscillation amplitude, in effect melting the emergent time periodicity.  

In Fig.~\ref{fig3_FiniteTemp}(a) we fix $N=3$ and examine the behavior of $\langle \hat\sigma_x^{(2)}\rangle$ for the same initial state, Eq.~\eqref{initialstate} and we consider several values for the inverse temperature, $\beta$. It is clear that $\beta=10$ is sufficiently large to recover the zero-temperature behavior. As we increase the temperature of the collisional bath, we find that the oscillations decay in time. While the decay appears slow for large values of $\beta=5$, it is nevertheless exponential, which is more clearly evident by examining smaller values of $\beta$. From the inset we see that for low, but non-zero temperatures, the oscillatory behavior can still be observed, albeit the frequency shows a slight drift from the value predicted by Eq.~\eqref{OscFreq}. 

The impact of temperature on the oscillation amplitude is captured in Fig.~\ref{fig3_FiniteTemp}(b) where we consider the ratio between $\langle \hat\sigma_x^{(2)} \rangle$ and $\langle \hat\sigma_x^{(2)} \rangle_{\beta\to\infty}$, i.e., the zero temperature value, at different instances in time for $t=200$ [left most black] to $t=500$ [rightmost lighter gray]. We see that at a fixed value of temperature, the oscillation amplitude monotonically decays with time, confirming the aforementioned melting of the dissipative time crystal for finite temperatures. Comparing the main panel for $N\!=\!3$ with the inset shows that the same qualitative behavior is found for $N\!=\!4$. However, we see that larger systems can delay, although not fully mitigate, the impact of the thermal environment since the onset of the decay is shifted to lower values of 
$\beta$. 

\subsection{GKSL master equation approach}
\begin{figure}[t]
(a)\\
\includegraphics[width=0.85\columnwidth]{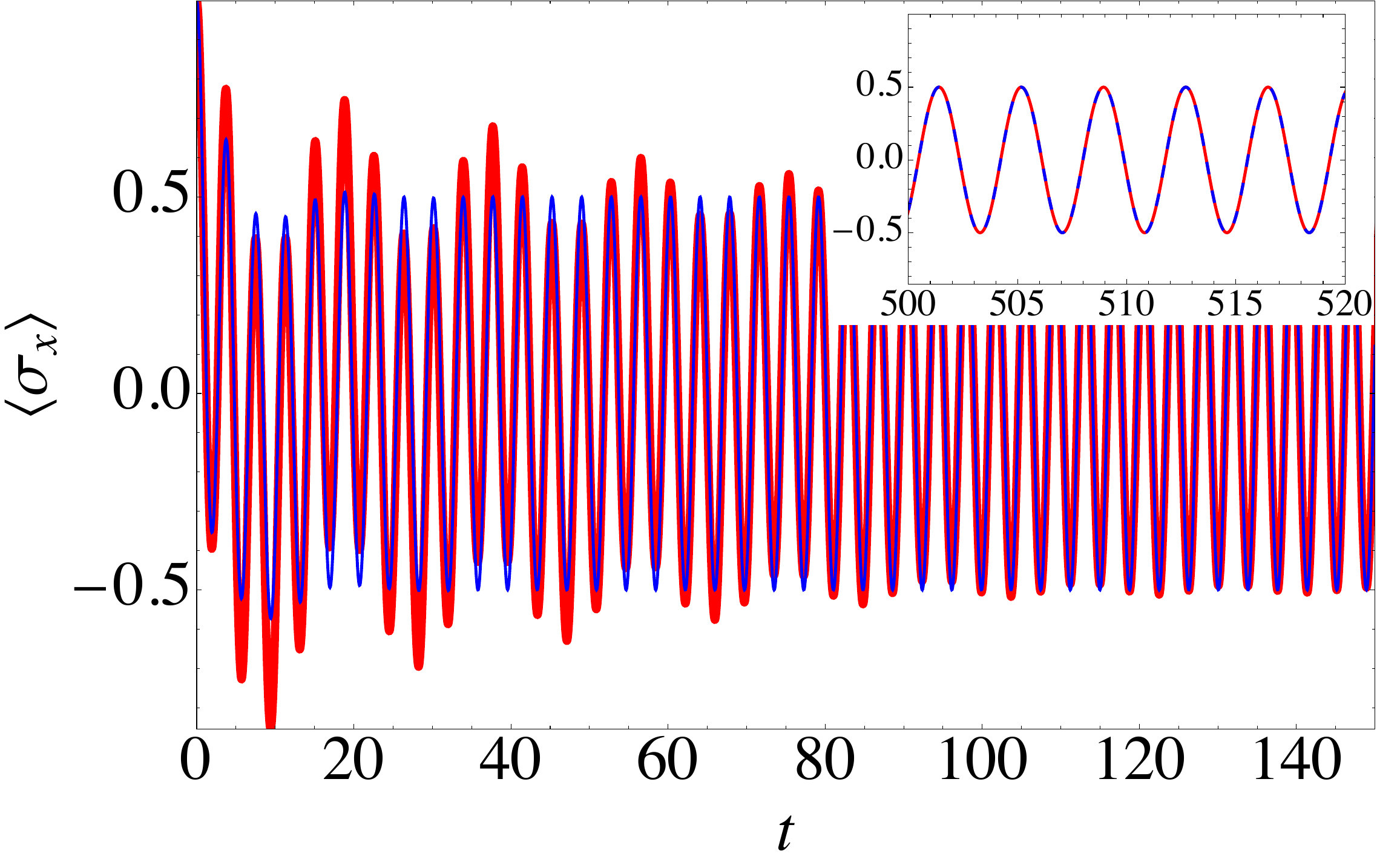}\\
(b)\\
\includegraphics[width=0.85\columnwidth]{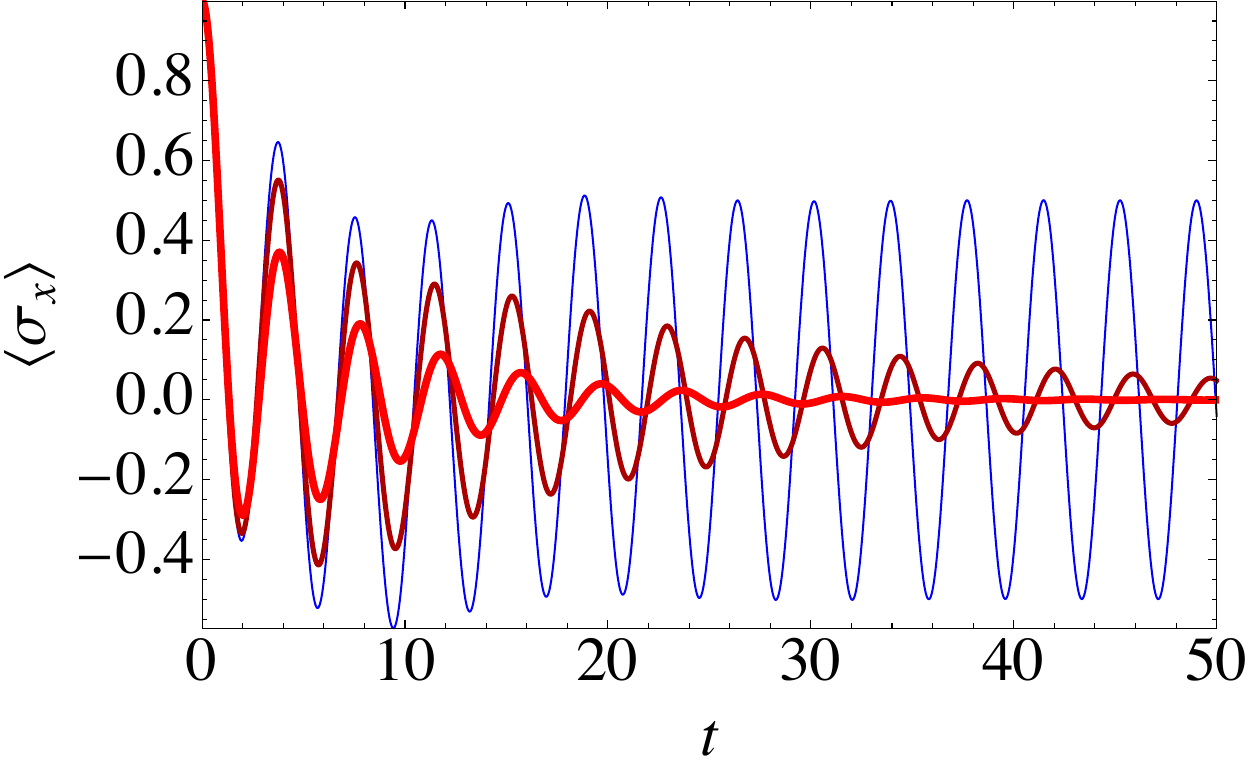}\\
(c)\\
\includegraphics[width=0.85\columnwidth]{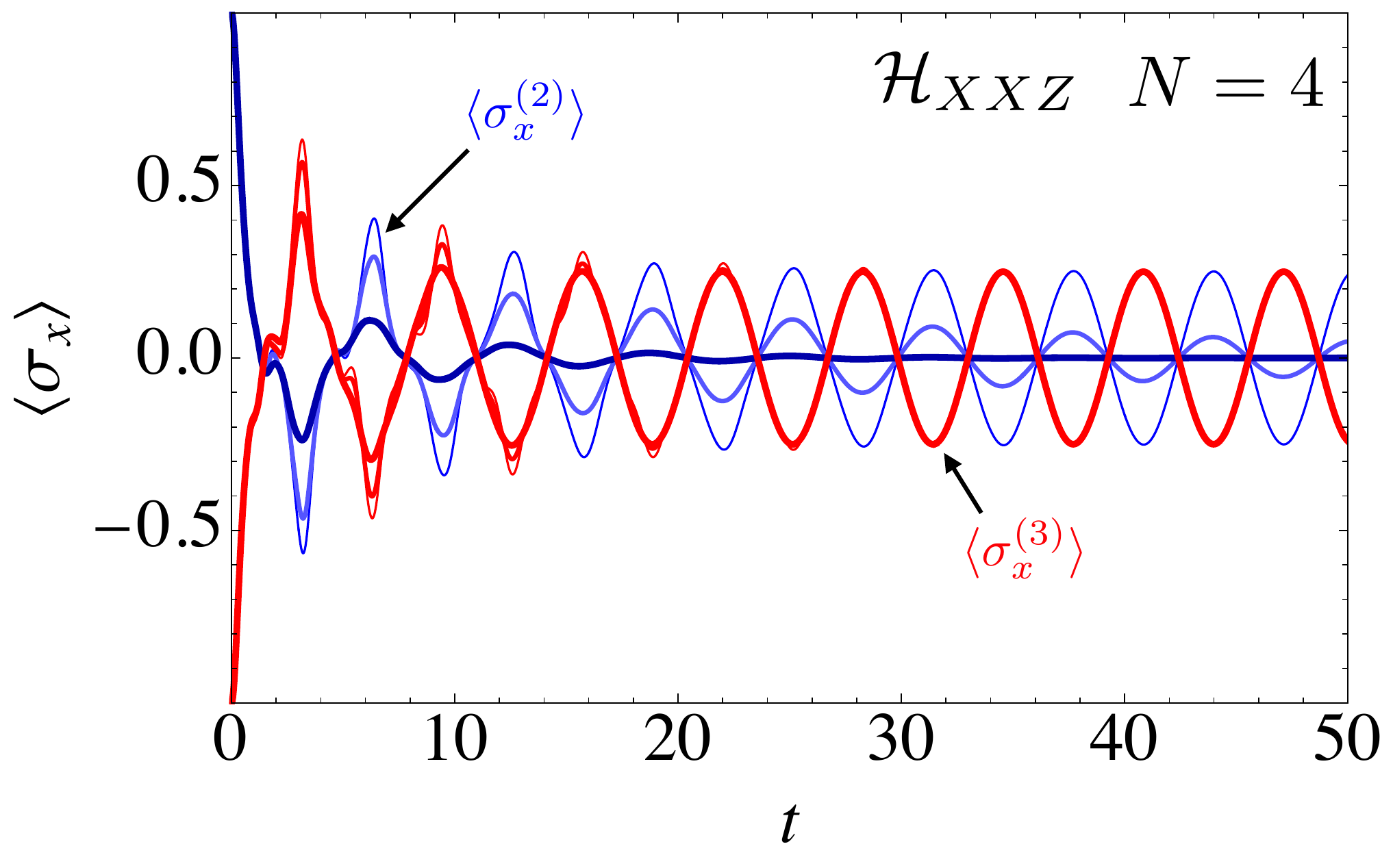}
\caption{(a) Collision model (thick, red) versus master equation simulation (thinner, blue) showing $\langle \sigma_x^{(2)} \rangle$ for $N=3$ with same initial state as before. For zero temperature, i.e., $\nbar=0$ in Eq.~\eqref{GKSL_ME} and $\beta=0$ in Eq.~\eqref{ancilla_state}. Inset shows a zoomed in region of time. (b) Master equation simulation for $N=3$ with same initial state as before, Eq.~\eqref{initialstate}, taking increasing values of temperature with $\nbar=0$ [thinnest, blue], 0.1 (dark, red), and 0.5 (lighter red, thickest). (c) $N=4$ for the XXZ model given by Eq.~\eqref{XX_Ham}. We show $\langle \sigma_x^{(2)} \rangle$ (blue) and $\langle \sigma_x^{(3)} \rangle$ (red) for different values of temperature $\nbar=0$, 0.1, and 0.5 (thinnest to thickest curves). The initial state  vector is $\ket{\Psi_0} = \ket{0}^{(1)} \otimes \ket{+}^{(2)} \otimes \ket{-}^{(3)}\otimes \ket{0}^{(4)}$. The oscillation frequency is $2B$ as expected from Ref.~\cite{GiacomoPRA}.}
\label{fig4_MasterEq}
\end{figure}

We now show that a consistent 
behavior can be observed by modelling the dynamics using the GKSL master equation
\begin{equation}
\label{GKSL_ME}
\begin{split}  
\dot {\hat\varrho} = & -i \left[ \hat{\mathcal{H}}, \hat\varrho \right] + \Gamma\left(\overline{n} + 1\right)\left(\hat\sigma_-^{\left(1\right)} \hat\varrho \hat\sigma_+^{\left(1\right)} - \frac{1}{2}\left\{\hat\sigma_+^{\left(1\right)}\hat\sigma_-^{\left(1\right)}, \hat\varrho \right\} \right) \\ &+ \Gamma\overline{n} \left(\hat\sigma_+^{\left(1\right)} \hat\varrho \hat\sigma_-^{\left(1\right)} - \frac{1}{2}\left\{\hat\sigma_-^{\left(1\right)}\hat\sigma_+^{\left(1\right)}, \hat\varrho \right\} \right)
\end{split}
\end{equation}
with $\Gamma>0$ being the damping rate, $\nbar=(e^{\beta B}-1)^{-1}$ is the thermal occupation number, and the bath is acting on only a single qubit in the system, i.e., $q_1$. 
A microscopic derivation of Eq.~\eqref{GKSL_ME} notoriously requires several approximations; commonly this can be obtained when system and bath are weakly coupled and the dynamics of the latter is much faster than any timescale of the system (Born-Markov, secular approximation)~\cite{Breuer2002}. However a different, less common microscopic derivation can be obtained within the collisional picture in the limit of short collision times~\cite{GiacomoPLA}, provided the interaction Hamiltonian Eq.~\eqref{interactionHam} is rescaled as $\tau^{-1/2}$ corresponding to the ultra-strong coupling regime~\cite{cMPS2,Ciccarello_2022,GiacomoPLA}. 
It is important to remark however that, since we are only interested in the long-time dynamics and not in the transient, we do not perform here such rescaling of the interaction to the ultra-strong coupling regime and treat the master equation Eq.~\eqref{GKSL_ME} as an \textit{effective} reduced description, that qualitatively matches in the steady-state regime of the numerical simulations with the collision model. In Fig.~\ref{fig4_MasterEq}(a), we show $\langle \hat\sigma_x^{(2)} \rangle$ for the random arrival time collision model (thicker, red) and from the corresponding master equation approach (thinner, blue) using Eq.~\eqref{GKSL_ME} for the zero temperature case, i.e., $\beta\to\infty$ in Eq.~\eqref{ancilla_state} and $\nbar\to0$ in Eq.~\eqref{GKSL_ME}, again taking the initial state to be Eq.~\eqref{initialstate}. During the transient we see that the qualitative behavior is consistent, although there are clear quantitative differences. However, once the system reaches its steady state, which corresponds to $q_1$ being driven into its local ground state vector $\ket{0}^{(1)}$, the inset shows that the dynamics in the two pictures perfectly overlap. In Fig.~\ref{fig4_MasterEq}(b) for finite temperature we observe a consistent (although not strictly identical) behavior when compared with the collision model approach at the same temperature [cfr.\ Fig.~\ref{fig3_FiniteTemp}(a)]. Having established that the dynamics is accurately captured by employing the master equation, this then allows us to more carefully assess whether the time-periodicity observed strictly corresponds to a dissipative time-crystal~\cite{BucaNatComms}.

\subsection{Correspondence with dissipative time crystals}
\label{BucaConds}

The melting of the periodic behavior exhibited for finite temperatures raises questions as to whether the observed behavior constitutes a bonafide time crystal. This can be readily checked by constructing the associated dynamical symmetry $\hat{\mathcal{A}}$ for the model and testing the criteria set out in Ref.~\cite{BucaNatComms}, where given the correspondence established between the collisional model and the GKSL master equations, we consider the jump operators in Eq.~\eqref{GKSL_ME} and check
\begin{equation}
\label{BucaConditions}
\begin{split}
&(i)~\left[\hat{\mathcal{H}}, \hat{\mathcal{A}} \right] \hat\rho_\infty = -\lambda \hat{\mathcal{A}} \hat\rho_\infty,~ \text{and},\\
&(ii)~\left[ \hat\sigma_k^{(1)} , \hat{\mathcal{A}} \right] \hat\rho_\infty = \left[ \left(\hat\sigma_k^{(1)}\right)^\dagger , \hat{\mathcal{A}} \right] \hat\sigma_k \hat\rho_\infty = 0
\end{split}
\end{equation}
for $k=\{+,-\}$ and where $\hat\rho_\infty$ is the steady state of the dynamics. For the $N\!=\!3$ case we can easily diagonalise Eq.~\eqref{LMG_Ham} and determine the eigenstates that support the observed oscillations from which we can obtain the dynamical symmetry. Considering the following state vectors
\begin{equation}
\label{LMG_dynsym}
\begin{split}
\ket{\psi} &= \ket{0}^{(1)}\ket{0}^{(2)}\ket{0}^{(3)},\\
\ket{\phi} &= \ket{0}^{(1)}\ket{1}^{(2)} \ket{0}^{(3)} -\ket{0}^{(1)} \ket{0}^{(2)} \ket{1}^{(3)},
\end{split}
\end{equation}
we find that $\mathcal{A}=\ketbra{\psi}{\phi}$ satisfies the conditions of being a dynamical symmetry {\it only} for the zero temperature bath, i.e., $\left[\mathcal{H}_S,\mathcal{A}\right]=-\left(\tfrac{2}{3}+2B\right)\hat{\mathcal{A}}$ and while condition {\it (ii)} is only valid for $k=-$. This therefore explains why the presence of finite temperatures impacts the dynamics since the symmetry is not protected and consequently the time-crystal melts. The crucial point is the state of $q_1$ appearing in Eq.~\eqref{LMG_dynsym}: in both $\ket{\psi}$ and $\ket{\phi}$, $q_1$ is in its local ground state, and since this is the state that the zero temperature bath drives it to, this ensures that once the system has purged all other unwanted eigenstates into the dissipative bath via $q_1$ the whole system will remain trapped in the decoherence free subspace. In contrast, finite temperatures drive $q_1$ to a thermal state with some population in the state vector $\ket{1}^{(1)}$ which ultimately ensures that the system can never ``settle" into the decoherence free subspace.

The above considerations gives cause to revisit the $XXZ$ spin ring model considered in Ref.~\cite{GiacomoPRA}
\begin{equation}
    \label{XX_Ham}
    \hat{\Ham} = J \sum_{i=1}^{N-1} \left( \hat{\sigma}^{(i)}_x \hat{\sigma}^{(i+1)}_x +\hat{\sigma}^{(i)}_y \hat{\sigma}^{(i+1)}_y \right) + B\sum_{i=1}^N \hat{\sigma}_z^{(i)},
\end{equation}
where we have omitted $\hat{\sigma}^{(i)}_z \hat{\sigma}^{(i+1)}_z$ term as it is immaterial to our discussion. In Ref.~\cite{GiacomoPRA} a qualitatively similar behavior as discussed thus far for Eq.~\eqref{LMG_Ham} 
has been demonstrated for this model, however, the collisional bath was assumed to be zero temperature. 
We can again use the GKSL master equation to examine the impact of temperature on the time-periodic behavior since the long time behavior is accurately captured. In Fig.~\ref{fig4_MasterEq}(c) we show $\langle \hat\sigma_x^{(2)} \rangle$ (blue) and $\langle \hat\sigma_x^{(3)} \rangle$ and consider several different temperatures corresponding to $\nbar=0$, 0.1, 
and 0.5. We see that $\langle \hat\sigma_x^{(2)} \rangle$ decays as a function $\nbar$. In contrast, $\langle \hat\sigma_x^{(3)} \rangle$ shows persistent oscillations regardless of the value of temperature, with the oscillation frequency given by $2B$ which is exactly the expected behavior for the dynamical symmetry given by
\begin{equation}
\label{GiacSymmetry}
\begin{split}
\hat{\mathcal{A}}_1&=1\!\!1^{(1)} \otimes \ketbra{\vartheta_1}{\varphi_1} , \\
\ket{\vartheta_1} &= \ket{0}^{(2)}\ket{0}^{(3)}\ket{1}^{(4)} - \ket{1}^{(2)}\ket{0}^{(3)}\ket{0}^{(4)} ,\\
\ket{\varphi_1} &= \ket{0}^{(2)} \ket{1}^{(3)} \ket{1}^{(4)} -\ket{1}^{(2)}\ket{1}^{(3)} \ket{0}^{(4)},
\end{split}
\end{equation}
as reported in Ref.~\cite{GiacomoPRA}. It is readily confirmed that $\hat{\mathcal{A}}_1$ satisfies all the conditions given by Eq.~\eqref{BucaConditions} for {\it both} jump operators. The fact that $q_3$ exhibits the oscillations can be seen from Eq.~\eqref{GiacSymmetry} as $q_3$ in the spin ring is the only state that differs between $\ket{\vartheta_1}$ and $\ket{\varphi_1}$. 
The resilience to thermal effects can be understood as a consequence of the fact that the state of $q_1$, which is in contact with the thermal bath, in $\hat{\mathcal{A}}_1$ is simply the identity, and, therefore, the bath can drive the qubit to any state and still allow for the emergence of stable time periodic behaviors. Only $q_3$ exhibits such features as can be confirmed by examining the behavior of $\langle\hat\sigma_x^{(2)}\rangle$ in Fig.~\ref{fig4_MasterEq}(c) where we see that introducing temperature leads to an asymptotic decay of the oscillation amplitudes. This is because the dynamical symmetry that is supported on sites $q_2$ and $q_4$ corresponds to
\begin{equation}
\label{GiacSymmetry2}
\begin{split}
\hat{\mathcal{A}}_2&= \ketbra{\vartheta_2}{\varphi_2} , \\
\ket{\vartheta_2} &= \ket{0}^{(1)}\ket{1}^{(2)}\ket{0}^{(3)}\ket{0}^{(4)} - \ket{0}^{(1)}\ket{0}^{(2)}\ket{0}^{(3)}\ket{1}^{(4)},\\
\ket{\varphi_2} &= \ket{0}^{(1)}\ket{0}^{(2)} \ket{0}^{(3)} \ket{0}^{(4)}.
\end{split}
\end{equation}
For $\hat{\mathcal{A}}_2$, similar to Eq.~\eqref{LMG_dynsym}, the state of $q_1$ is restricted to be  $\ket{0}^{(1)}$. The necessity for $q_1$ to remain in its ground state means this symmetry can only be maintained for zero temperature environments. This can be further confirmed by computing Eq.~\eqref{BucaConditions} for $\hat{\mathcal{A}}_2$ where again we find that condition {\it (ii)} is only satisfied for $k=-$. We can therefore conclude that a the emergent time periodicity associated to $\hat{\mathcal{A}}_1$ for a 4 qubit system with XX(Z) type interactions is remarkably robust and does not melt in the presence of finite temperatures.

Our setup, according to the definition and formalism put forward in Refs.~\cite{BucaNatComms, BucaNJP2020, BucaSciPost2022}, can thus be classified as a dissipative time-crystal due to the emergent oscillatory behavior in system observables. As we have seen, there is a transient that the system must undergo where the bath ``purges'' eigenstates that do not correspond to the dark states of the Hamiltonian, and, therefore, the timescales necessary for a system to exhibit such time-periodic behavior is largely governed by the Liouvillian gap~\cite{BucaNJP2020} or that of the iterated quantum channel reflecting the collisional dynamics. We have demonstrated that the robustness of dissipative time-crystals is delicately dependent on the structure of the dynamical symmetry governing the oscillations. Our results indicate that characterising emergent oscillations as a bonafide time-crystal should be done with caution, with a careful analysis of the robustness of any such features, e.g., as considered here or in the thermodynamic limit along the lines of Ref.~\cite{riera2020time}, being prudent.

\section{Conclusions}
We have examined the stability of emergent time-periodicity in an interacting system modelled by the Lipkin-Meshkov-Glick Hamiltonian. By coupling one of the constituent qubits of the system to a collisional bath, we have shown that a stable time-periodic phase can be established when the bath is at zero temperature. Introducing finite temperatures leads to a melting away of the time periodicity and we have shown that this is due to the characteristics of the eigenstates that define the decoherence free subspace corresponding to the relevant generalised dynamical symmetry~\cite{BucaNatComms}. We have shown that the nature of the microscopic interactions within the system play a vital role in the stability of such dissipative time-crystals, showing that by changing interaction model to the nearest neighbor $XXZ$ type can allow for a thermally robust emergent time-periodicity. 

Our results therefore provide useful insight into the conditions necessary to realise stable time-periodic behavior and leave open several relevant directions for future study; for example considering the impact that non-Markovian dynamics~\cite{CampbellnonMarkovCMs} or environmental coherence~\cite{LandiCoherentCMs} can have on the steady-state behavior. While for non-Markovian dynamics, provided the fixed point of the generator drives the qubit in contact with the bath to a state compatible with the emergent time-periodic behavior, we expect a qualitatively similar behavior as discussed here to be established, the impact of coherence in the incoming environmental auxiliary systems' states is less evident. This work can also be seen as a further invitation to take steps towards probing time-periodic behaviour under controlled conditions using instances of quantum simulators \cite{KhemaniPRXQ, Augustine2DTC,Zheng2024arXiv}. It is the hope that the present work contributes to stimulating such steps.

\acknowledgements
SC is grateful to Brian Creed and Eoin O'Connor for useful discussions. SC acknowledges support from the John Templeton Foundation Grant ID 62422 and the Alexander von Humboldt Foundation. JE has been supported by the DFG (FOR 2724, CRC 183), the BMBF (MUNIQC-Atoms) and the Quantum Flagship (PasQuanS2). GG acknowledges financial support from the Rita-Levi Montalcini scheme.
\bibliography{references_timecrystals}

\end{document}